RESEARCH ARTICLE                                                                                                          OPEN ACCESS

# Comparative Analysis of Deep Learning Algorithms for Classification of COVID-19 X-Ray Images

Unsa Maheen [1], Khawar Iqbal Malik [2], Gohar Ali [3]

[1], [2], [3] Department of Computer Science, University of Lahore – Pakistan

**ABSTRACT**

The Coronavirus was first emerged in December, in the city of China named Wuhan in 2019 and spread quickly all over the world. It has very harmful effects all over the global economy, education, social, daily living and general health of humans. To restrict the quick expansion of the disease initially, main difficulty is to explore the positive corona patients as quickly as possible. As there are no automatic tool kits accessible the requirement for supplementary diagnostic tools has risen up. Previous studies have findings acquired from radiological techniques proposed that this kind of images have important details related to the coronavirus. The usage of modified Artificial Intelligence (AI) system in combination with radio-graphical images can be fruitful for the precise and exact solution of this virus and can also be helpful to conquer the issue of deficiency of professional physicians in distant villages. In our research, we analyze the different techniques for the detection of COVID-19 using X-Ray radiographic images of the chest, we examined the different pre-trained CNN models AlexNet, VGG-16, MobileNet-V2, SqeezeNet, ResNet-34, ResNet-50 and COVIDX-Net to correct analytics for classification system of COVID-19. Our study shows that the pre trained CNN Model with ResNet-34 technique gives the higher accuracy rate of 98.33, 96.77% precision, and 98.36 F1-score, which is better than other CNN techniques. Our model may be helpful for the researchers to fine train the CNN model for the the quick screening of COVID patients.

*Keywords:* - COVID-19, X-Ray, classification, Image, CNN,

## I. INTRODUCTION

COVID is responsible for the deaths of approximately 1.94M patients suffering with COVID in 2019-2021 making it the leading cause of COVID Virus related mortality. More than, 90.4M individuals worldwide are considered to be infected with COVID Virus. For the diagnosis of COVID, chest radiography is a common and inexpensive method of quickly examining the lungs and chest condition. Yet, the COVID Virus was identified when the overall symptomatic case fertility risk (the chance of dying after acquiring symptoms) of COVID in Wuhan was 1.4 percent (0.9 percent – 2.1 percent). Since COVID is both treatable by using the vaccines and often fatal in patients with low immunity, improving its diagnosis is of the utmost importance. During writing this paper the accumulative number of diagnosis worldwide has increased 90,82,986, which seriously harmful for the lives and the health of the humans all over the world. COVID is the caused by type of virus called Severe Acute Respiratory Syndrome COVID Virus 2 (SARS-CoV-2) [1]. The biggest problem at present is the recognizing and investigating of COVID. This type of virus can be detected by using the RT-PCR (Real Time Reverse Transcriptase Polymerase Chain Reaction) for diagnosing viral nucleic acids as a baseline standard [2].

Now, COVID virus outbreak all over the world, the number of mortality and morbidity rate are rising day by day according to the updated reports of the World Health Organization (WHO). However, due to a huge range of epidemics, it is not enough to test all the infected individuals in high incident areas and countries and also it is impossible to do RT-PCR tests on Thousands of infected patients not only this diagnosing way to takes many hours or even days to be finished [3].

On 11th February 2020, WHO Director-General send acronym "COVID". This infection is formed by a severe virus is called COVID virus. From recent years, two COVID virus types are finding, i.e. SARS-CoV and MERS-CoV. It is started in china, spread to twenty-four different countries rapidly and 8000 cases & 800 of deaths recorded. Then after this it started in Saudi Arabia, here 2500 cases and 800 deaths recorded. Usually, approximately 2% of the total strength of healthy carriers of a CoV, these viruses are calculated about 5% to 10% of lung infections. [4] it takes approximately 14 days to show the different symptoms of this harmful disease in the infected individual, still there were no proper treatment, medicine or vaccine or drug is present to fight against this severe virus. Medical imaging technique play a vital role of x-rays and computed tomography (CT) for testing infected people of COVID disease.

Researchers [5] observed Chest scanning pictures like x-rays computed tomography scans (CT scan), MRI (magnetic resonance imaging) have been used for investigating the morphological patterns of the lesions of lungs related to COVID [3]. However, the perfection of detecting COVID by Chest scanning is deeply depends upon the specialists and deep learning approaches were investigated as a tool for brutalising and diagnosing. Medical professionals can use chest photographs to recognise the architecture of formations of the chest and to see their forms, sizes, densities, and qualities. Due to these considerations, a variety of deep learning-based approaches for coronavirus have been proposed. [6].

A serious stage is here an actual and correct showing of the COVID patients not only getting rapid treatment and isolation after the public to stop distribution of the viral infection as well. Serology and inverse record of polymerase chain reaction, i.e., rRT-PCR, are state-of-the-art procedures for identifying COVID. Serology for antibody recognition is a service that provides scientific systems and people observation. Because the testing kits aren't universally available, it's fascinating to observe who's been infected by the virus. Moreover, possible trials consume sometime to give the results, time overwhelming sometime fault disposed to in the existing state of emergency. So, a quicker and consistent showing methods that might be more established by the PCR test is directly essential. Several studies have used radiography like X-rays as well as computed tomography (CT-scans) could determine the new COVID virus's distinctive symptoms. [7]

New research suggests that chest radiography is being used for early COVID testing in outbreak regions. (Aminian, 2020). As either a result, radiographic image testing can be employed as an alternative towards the PCR approach, though in some instances has a greater sensitivity [8].

In the genuine assessment of medical pictures, deep learning (DL) detection architectures have demonstrated outstanding effectiveness. The categorization of pictures with highly distinct features is a commonly used system of deep learning-based visual evaluation. Deep learning-based approaches are extensible, possible to automate, and simple to use in medical practice [9].

To correctly allocate input pictures in the training instances, this approach uses picture sectionalizing, recognition of efficient picture characteristics derived from sectioned zone inside temporal, and the creation of optimum Deep learning based prediction architecture [10]. The researcher utilised a Convolutional Neural Network (CNN) to create an architecture with an optimal set of synthesized attributes which can differentiate coronavirus disease pictures from non-coronavirus disease instances with an accuracy of 86%. [11].

In advance, modern research mostly depends artificial intelligence (AI) in which includes data science and deep learning (DL). These methods are used for stop spreading of COVID, correct diagnosis of disease, precautionary measures drugs and vaccines discovery, proper in time treatment and much more [12]. Deep Learning (DL) has required big dataset with more powerful resources of implementation. Small dataset is normally common for detecting pandemic disease Deep Transfer Learning (DTL) is more powerfully effective for detecting and learning different tasks on another side moreover, edge devices, internet of thinking (IOT), robots, drones are very effective in this severe situation these devices provides infrastructure for automated detection of COVID in critical condition. Deep Transfer Learning (DTL) is most useful than Edge computing which provides the pipeline for mitigation of any severe acute disease [13].

Convolutional neural networks (CNNs) are among the most well-known deep learning architectures. The innovation in CNNs came in race where the error rate was approximately halved for object identification [14]. The suggested network is operated as a visual interface on an input system with either a camera then it produces chest image categorization with much less than 1 second. By use of convolutional neural networks (CNNs) to evaluate medical pictures in order to give computer-aided diagnosis has piqued attention (CAD). According to current research, image classification CNNs may still not extend to it to fresh data as initially assumed. For a synthetic pneumonia diagnostic job, we evaluated how effectively CNNs extended over three healthcare organizations. [15]. Deep learning (DL) suggest quick, efficient strategies to find deformity and to find actual features of altered lung parenchyma, that may be link with COVID virus. However, the present symptomatic COVID datasets on different architectures are not enough to describe deep neural networks. We will use COVID chest pictures datasets to describe best deep learning techniques for better results. Deep neural networks (DNNs) includes at biggest level for pixels wise classification. Deep Learning performed high performance in medical. However, different ways for disease perception on refining the precision of estimates. In computer-based medical identification is compulsory for obtaining trust in field. In our research, we analyze the different techniques for the recognition of COVID-19 using X-Ray radiographic images of the chest, we examined the different previously learned CNN techniques VGG-16, SqeezeNet, ResNet-34, AlexNet, ResNet-50, MobileNet-V2, and COVIDX-Net to correct analytics for classification of COVID-19.

### A. Problem Statement

As now a days COVID symptomatic patient reports take 2-3 days approximately 48-72 hours for confirmation, by using chest image classification immediately report COVID patient to save other humans. Here, I have studied different methods on chest pictures and observed various results of chest pictures classification on different diseases, now by apply deep learning by using CNNs (Convolutional Neural Network) on COVID chest pictures to find the reason of COVID and how it is different from other diseases and we will compare COVID patient to other diseases of patients by using deep learning, however during previous study of proposed work we have found good performance of CNN on other diseases of radiography (diagnostic and therapeutic).

Moreover, when we compare CNN models on COVID disease, for better performance.

### *B.  Study Background*

Applying deep learning methods, this work aimed to develop an advance recognition framework to describe COVID among IAVP & fit patients by using lung CT scans. A number of 618 CT item analysis were discovered: 219 sample were collected from 110 COVID patients of different ages. The Noisy-OR Bayesian method was used to determine the illness category and cumulative ranking of a CT scan case. The entire result reached 86.7 percent of all CT scan patients combined, according to the outcomes ratio of the dataset consisting. The research for the treatment of COVID victims was recognized by machine learning techniques. [5]. This deep learning-based method for recognizing and detecting pneumonia within chest X-rays (CXRs). CXRs were frequently utilized by researchers to identify imaging studies. For pixel-by-pixel classification, a deep convolutional neural network that integrates global information is used. This architecture obtains better results to evaluated on chest radiograph datasets, which represents possible pneumonia origin [16]. ANN is completely like a human brain or we can say that like human nervous system having so many connected nodes. Each node has some state and positive or negative weight that can be used to activate or deactivate the node. ANN is helpful in training the machine using sample data. The trained machine is used to detect the pattern of hidden data.

Its approach uses picture sectionalizing, recognition of operative pictures characteristics was created using a sectioned region throughout the temporal, as well as an optimal Deep learning based prediction architecture to properly identify source images to training sets. It utilized a Convolutional Neural Network (CNN) to produce a framework by using optimal collection of synthesized attributes which differentiate coronavirus disease pictures from non-COVID instances with an efficiency of 86% [17]. Deep learning technique is efficiently used to find diseases and brings out the text-based features of the COVID virus. Based on convolutional neural networks open standard chest radiological pictures for COVID sufferers have been offered for this reason. In comparison to other approaches or algorithms, machine learning outcomes provide more consistency in findings.

## II.  LITERATURE REVIEW & RELATED WORK

### *A.  COVIDX-Net:*

Coronaviruses are the primary reason of severe acute respiratory syndrome (SARS-CoV) & Middle East Respiratory Syndrome coronavirus (MERS-CoV). This coronavirus infection was first found in Wuhan, China, in 2019. It has since spread around the world, affecting many individuals and resulting in fatalities. As a result, the primary aim is to establish COVIDX-Net, an image recognition framework that assists radiologists in evaluating Covid-19 using a variety of X-Ray visualization was 86.7 percent [18].

This inception V3 algorithm had the lowest classification results, with F1 values of 0.67 percent for usual circumstances while 0.60 for coronavirus instances. COVIDX-Net has been especially focused on eight percent and twenty percent of X-Ray pictures acquired throughout the system learning & validation stages. With F1 scores 0.89, 0.91 for normal individuals and COVID-19, either VGG19 and deep convolutional architectures show higher and similar efficiency in automation coronavirus identification. Deep neural learning algorithm is useful to classify CoV in X-Rays visualization depends upon suggested COVIDX-Net architecture.

### *B.  Automatic Detection of Coronavirus Disease (COVID-19)*

In 2019 coronavirus start from China & spread everywhere in the world, in comparison to the continuously growing amount of COVID-19 kits accessible in hospitals, there are a restricted amount of COVID-19 kits present in health care centres. A globe's quantity of medical professionals per people is obviously minimal. To protect individuals from contracting COVID-19, an autonomous detector must be used different ResNet and Inception models are numerous early learning deep convolutional system algorithms for coronavirus identification utilizing X-ray tomography. By employing four categories fivefold cross validation, these three distinct binary categorizations involving four classes. The best output for dataset-1 is 96.1 percent, 99.5 percent for dataset-2, and 99.7 percent for dataset-3, according to the actual outcomes of its pre-trained ResNet50 algorithm [19]

341 number of CoV victims' tomographic images of chest obtained through GitHub. The disease affected the lungs badly and tomographic images tells the clear image of lungs, so that is better representation to detect the virus from visual radiographic pictures. ChestX-ray8 produced 2800 clear (healthful) chest radiographic X-ray pictures. In the analyses, all pictures were scaled to 224x224 pixels.

Deep neural systems design subfield of artificial intelligence which were built on the human brain organization. Consisting realm of clinical photo analysing, deep learning algorithms currently provide very reliable findings. Throughout the medical area, deep learning algorithms produce perfectly alright outcomes. The CNNs is the example of DNN which is used to solve image processing problems and that is very useful for the classification of image. It performs rapidly and effectively, CNN has 3 layers: a convolutional neural system, a max - pooling, and a fully - connected layers. The feature extraction process used first two layers of CNN convolutional and pooling layers. By using these two layers the classification process gives results in the form of fully connected layer.

COVID-19 sufferers were recently identified, preventing the infection from occurring to all others. For the identification of COVID-19 sufferers, a neural transfer learning-based technique employing X-ray pictures of normal, affected with corona, bacterial disease & influenza victims were presented. In five models using three distinct datasets, the effectiveness of the ResNet50 pre-trained classifier is the best. Due to the extreme outstanding performance in findings, it appears to us that it'll be useful to physicians in medical practice. In detection of coronavirus disease, recently their research will provide how deep transfer learning methods are useful for finding accurate results rapidly. On the other hand, the CNNs model's classification findings may be utilized to assess a growing list of COVID19 tomographic x-rays pictures.

### C. *Detection of coronavirus based on Deep Features and SVM*

The recognition of coronavirus disease in patient is sensitive task in clinical field now a day. This virus is spreading speedily between people approximately 100,000 people all over the world. In this situation, it is necessary to detect the infected people for prevention to spread. The rich element of a support vector machine (SVM)-based approach is utilized to detect coronavirus infections utilizing X-ray pictures. SVM is utilized for categorization instead of a deep learning-based predictor, and a large dataset is employed for training and test data. For identification, deep features from the CNN model's fully - connected convolutional layers are employed, as well as SVM. The SVM is used to categorize corona-affected X-ray pictures. COVID-19 infection, bacterial pneumonia and normal individual Xray images are included.

The rich characteristics of Classification algorithms are utilized to discover COVID-19 using SVM. Employing ResNet50's deep characteristic, the SVM provides the best efficiency. For the identification of COVID-19, the proposed framework, ResNet50 with SVM, obtained an accuracy, specificity, FPR, and F1 scores of 95.33 percent, 95.33 percent, 2.33 percent, and 95.34 percent, respectively (ignoring SARS, MERS and ARDS). ResNet50 Combined SVM achieves the greatest result of 98.66 percent. The effectiveness is dependent on the freeware X-ray pictures accessible on GitHub and Kaggle. The dataset contains a large amount of data, and the categorization is based on SVM. The classifying technique's assessment has been completed. LBP with SVM obtained 93.4 percent accuracy in a classic picture classification technique. Coronavirus, COVID-19, diagnostic, deep features, and SVM are utilized as keywords.

Proposed method makes advantage of CNN systems extracted CNN. All classifying algorithms' classification evaluation is performed. By using a Classification model, the COVID-19 illness is classified based on deep characteristics. A CNN is just a layered structure system in which every layer responds. The key service is provided by the layers, which then move to the following layer. The action is done in the GPU with a 64-bit micro batch size and enough GPU RAM to store the picture dataset. The activated product is a table, and the mechanism often used train the SVM. This method gives the entire multiclass error free results that have been learned. The fitcecoc algorithm employs the binary SVM classifier K(K-1)/2 and the 1 Versus All coding scheme. In which K is a one-of-a-kind classification designation [20].

The usability testing a One-Vs-All method with a straight SVM classifier. The effectiveness is determined by the number of iterations (20). The validity of several evaluation metrics with its mean, minimal, and maximal determined in 20 separate iterations has an 80:20 training / testing proportion. The coronavirus findings are based on information supplied by the WHO, the European Institute for Prevention Of illness, an organization of the EU, as well as other webstores throughout the worldwide. They collect chest X-ray pictures from free source GitHub and Kaggle for dynamic simulations. Deep feature and SVM are helpful methodologies for identifying coronavirus (COVOD-19) utilizing X-ray pictures. The deep characteristic of 13 previously trained CNN techniques to classifiers separately were used. Identification model was run 20 times on each side output was obtained. In comparison towards the other 12 classification methods, the ResNet50 + SVM recognition system demonstrated effectiveness. The suggested categorization algorithm for identifying COVID-19 automated has a 95.33 percent high accuracy. The overall result of 20 repetitions is 95.33 percent, while the accuracy performance attained is 98.66 percent. This study is also applicable to big datasets. When a victim is in a life-threatening condition and unable to attend for an X-ray, this technique comes in handy.

### D. *COVID-19 detection using deep learning models to exploit Social Mimic Optimization*

Coronavirus origins an extensive variability of lung infections and type of RNA virus that infects all living things human or animals. It is also a source of pneumonia in human beings. Artificial intelligence is useful for analyses in medical field. Here Coronavirus find-by using deep learning models, that is part of artificial intelligence. Here dataset contains three classes namely that are coronavirus, pneumonia, and normal X-ray images. Here the data classes use the Fuzzy Color technique as a pre-processing and the original images is arranged. Then datasets trained by using deep learning models (MobileNetV2, SqueezeNet) and the feature extracted after pre-processing and feature extraction by using Social Mimic optimization method. Afterward, effective features were joint and confidential used Support Vector Machines (SVM). Result is gathered with the planned approach was 99.27%. By using the given model, it is obvious that the model can competently subsidies to the discovery of COVID-19 virus.

MobileNet is a computational intelligence architecture with low system development. The MobileNet approach was used to accomplish object recognition, segmentation, and classification. The MobileNet framework, also known as MobileNetV1, MobileNetV2, is created by combining MobileNetV1 and MobileNetV2. The regularity among the layers is assessed by examining the MobileNetV2 framework to the previous. The sample variance for the MobileNetV2 network is 224 224 pixels, and the architecture in-depth (DW) is separated by filters in stages. The DW production has been improved, and the source characteristics have been split into two layers. Each component is divided into new layers, which are then joined for final output results. Till the procedure is done, each component is split into the next layer by merging it with the output characteristics. The ReLU tool was utilized among the stages in the MobileNetV2 architecture. To get to the simple phase, this classifier used the training date. On the output pictures, convolutional layer analysers are utilized to create the activation function. The input pictures are used to activate the function, which then moves to the next layer. For reliable findings, MobileNetV2 employed a pooling layer.

The above study uses 30% of the set of data for testing and 70% of the data pre-processing. The stacked dataset was cross-validated using the k-fold technique. SqueezeNet and MobileNetV2 features were utilized in the SVM technique classification. The training image is utilized to train the first stage CNN architecture, as well as the Fuzzy Color approach and the layered dataset for the SVM algorithm. In the categorization of the entire data, the SqueezeNet system had a total success rate of 84.56 percent. The dataset organized and used the Fuzzy Color method was categorized with a 95.58 percent optimum threshold rate in the second stage to use the SqueezeNet algorithm. The two phases of the Fuzzy Color method, on the other hand, assisted to the formation of the SqueezeNet architecture. The layered dataset was generated with 97.06 percent classification performance in the stage 3, which was accomplished with the SqueezeNet algorithm (overall accuracy) [21].

Infected people of COVID-19 mostly infected their lungs which cause the death. Here in this study a comparison of infected lungs with normal or pneumonia infected lungs which is not infected with COVID-19. COVID-19 detected via deep learning classification, moreover it is necessary the detection of COVID-19 which is expended very quickly in the worldwide. Artificial intelligence techniques used to find COVID-19 immediately and accurately. The novel approach is pre-processing steps on images is accurately feature extraction. In stack technique each pixel of image is superimposed which increases the efficiency of low pixel images. Here in this approach features are extracted accurately by using SMO algorithm. The mission of this model to obtain more accurate and quick results. By using the SMO algorithm, it improves the combined performance of classification. We use this model on smart devices by using MobileNetV2 model and also determine by using the mobile devices in spite hospital devices.by using this achieve maximum results in the cataloging of COVID-19 data 99.27 accomplishment obtain in the organization of healthy and pneumonia victim pictures, which are very good results.

*E. COVID-ResNet:*

As we know that COVID-19 is blowout in the world. To detect the variation COVID-19 to other diseases here is a study on chest x-ray pictures of patients that suffering in corona or some have other type of diseases like pneumonia.so here is a model to search this virus deeply here is a goal to identify this severe disease and open source dataset is present to identify this severe virus to other diseases accurately and effectively by using CNN (Convolutional Neural Network). Here's the detailed work that shows how to train residual neural networks quickly and effectively using progressive enlargement, cycle learning rate results, and discriminating learning charges. The outcome of this approach is a dataset that is readily accessible. This study is divided into three phases to offer fine-tuned appropriate results using a pre-trained ResNet-50 layout, which we call COVID-ResNet. It remains obtained gradually changing the size of input images 128x128x3,224x224x3,229x229x3pixels and accurate result of this network at every level. This method lengthwise automatic machine learning method gives fine-tuned efficient results 96.23%on COVIDx dataset with 41epochs. This approach provides a very affective and productive approach for multiple classes categorization of various forms of illnesses in healthy people. It provides help to find COVID-19 cases earlier without wasting time and overcome the burden in medical field. [22]

For classification purpose we use the pooling of CNN (Convolutional Neural Network) which gives the outstanding results in classification tasks. We apply different residual neural network on 50 layers is called ResNet50 (He). This model gives outstanding results and provide faster and quicker results of training data. On other hand the using factor of residual neural network design have ability to produce the trained data. This is important and sensitive part of training methodology that implies quicker faster and highly efficient results with some number if epochs by using this method introduce fastai (learning, 2020). By using COVIDx dataset rescale the images 224x224x3.these images are further reducing 128x128x128 and 229x229x3 and apply on different stages of training (subsection). The developers down sampled all pictures in the COVIDx collection to 224x224x3. These pictures are then resized to 128x128x3, 224x224x3, & 299x299x3 and used in training methods phases (Subsection D). The standard deviations of the pictures in the Neural network are used to adjust these photos. This seems to be risky because the network's parameters have been pre-trained using the ImageNet range of data. The learned method component is updated with some other set that includes a series of average/max pooling, batch normalization dropped off, or linearity segments for the transmission learning experience. [23] Now we see the performance of network and quantitative matrix. Including all sorts of infection and regular healthy patients, we employ specificity (recall), good predictive finesse (standardized uptake values), & F1-score. The estimated number of system parameters (both trainable and non-trainable), or even the overall number of input iterations and method efficiency on an alternative training dataset, as indicated by the researcher in his research.

COVID-ResNet is used here for the corona categorization with three contagious categories. COVID-ResNet initially accomplished using a freely accessible set of data, and COVIDx, as well as an alternative validation accuracy, and it performed exceptionally well throughout all classes. We further discuss the importance of feature extraction in terms of expanding the number of training samples collection and improving applicability. Through combining cutting-edge technology with human involvement during training, it is possible to find the best learning rates for boosting accuracy and scalability. Unfluctuating nevertheless COVID-ResNet is highly auspicious and precise, we also want to point out that it cannot be utilized independently aimed at medical finding. Main key area of research is demonstration that by combining several approaches, we may train networks which are much more conceptually successful. COVID-ResNet must be trained with a vast dataset and assessed in the outdoors using a bigger randomized controlled trial before it can be used for medical assessment.

### F. COVID-Net:

Coronavirus has a negative impact on the health of the community members, but the most important step in combating it is a thorough examination of reported cases, which itself is accomplished by radiological examination using chest radiographic. Initially, it was discovered that infections associated with COVID-19 have deformities in their chest radiographic images. Presently, inspired by all this public survey, COVID-Net, a fully convolutional neural architecture meant for monitoring cases of corona through X-Ray of Chest photographs of infected individuals that are readily accessible to all community residents, has been presented. COVID-Net seems to be a free software platform which noticing corona in Chest tomographs pictures during early stages an influenza. COVIDX is a publicly available comparative dataset that includes 13975 chest X-rays from 13870 individuals. In this project, COVID-19 instances are recognized from chest X-Ray pictures utilizing a human-machine combined design method, in which a human-driven concept system architecture design is combined with processor screening tool to produce a network topology. For either a faster convergence effectiveness and the tendency of selecting the most suitable, we will conduct both descriptive and inferential analysis to assess COVID-efficiency. Net's COVID-output, Net's together with the free and open - source feature and specification of open - source software input data building projects, is expected to be very useful, and it will foster the growth of a more accurate and consistent and indeed realistic computational intelligence solution for the assessment of COVID-19 using chest X-Ray images of affected individuals, as well as improve treatment of all those sufferers whom is diseased. [22]

### G. COVID Diagnosis-Net:

The critical action in the competition against CoV is to take influential examination of the site forming patients diagnosed with maximum of the prediction based on the monitoring of the genetic material of COVID-19 but the poor detection system is present with time taking operation but, in this process, radio graphical imaging is preferable in which chest X-Rays are used for the diagnosis. There are many studies which are using Deep Learning based solutions to examine the COVID-19 for chest X-Rays. In this study an artificial intelligence-based structure has been demonstrated for performing the existing studies the SqeezeNet with its light weight network design is tune up for COVID-19 diagnosis with Bayesian Optimization additive fine-tuned hyper parameters and augmented datasets make the purpose network to work much better than present network design and for obtaining a high quality and more accurate COVID-19 diagnosis.

CNN models are more accurate in image classification because of its self-learning qualities and superior classification results on many problems. CNN occurs when the layers of CNN is used with the rectified linear unit activation functions batch normalization operation and pooling layers (pool) for detection the COVID-19. [24] Bayesian Optimization gives the highly accurately outputs to make the propose network to work much better than the other models for obtaining the fine-tune results.

## III. RESULTS AND DISCUSSION

This section represents the output obtained from different experiments, here we are performing different analysis for detection of COVID-19 from Chest X-Ray images by using different pre-trained CNN models AlexNet, VGG-16, MobileNet-V2, SqeezeNet, ResNet-34, ResNet-50 and COVIDX-Net. We have analysed different models and obtain the comparative study on the basis of CNN models and find the best performing fine-tuned detection Model and also shown the comparison state-of-the-art results of different models. CNN models are used for evaluating Chest X-Ray images. The augmented samples of Chest X-Ray images used for training the models, after training data is divided into training and validation data by ratio of 70% and 30%. The validation result is used for prevent the model from overfitting and obtain the optimal results. The input X-Ray images were initially resized 224x224 while using AlexNet, VGG-16, MobileNet-V2, SqueezeNet, ResNet-34 and ResNet-50. Each model was trained for 50 epochs. The batch size and number of epochs resolute analytically. The performances of each model were determined depends upon different matrices that are F1-score, sensitivity, precision, specificity and accuracy. These matrices are computed by different parameters of confusion matrix I-e True-Positive (TP), True-Negative (TN), False-Positive (FP), False-Negative (FN).

TABLE I
COMPARISON OF NETOWRKS BY USISNG DIFFERENT PARAMETER

| Model | Precision (%) | Sensitivity (%) | Specificity (%) | F1 score | Accuracy (%) | AUC |
|---|---|---|---|---|---|---|
| ResNet-34 | 96.77 | 100.00 | 96.67 | 98.36 | 98.33 | 0.9836 |
| ResNet-50 | 95.24 | 100.00 | 95.00 | 97.56 | 97.50 | 0.9731 |
| GoogleNet | 96.67 | 96.67 | 96.67 | 96.67 | 96.67 | 0.9696 |
| VGG-16 | 95.08 | 96.67 | 95.00 | 95.87 | 95.83 | 0.9487 |
| AlexNet | 96.72 | 98.33 | 96.67 | 97.52 | 97.50 | 0.9642 |
| MobileNet-V2 | 98.24 | 93.33 | 98.33 | 95.73 | 95.83 | 0.9506 |
| Inception-V3 | 96.36 | 88.33 | 96.67 | 92.17 | 92.50 | 0.9342 |

### A. Results

The training performance is determined by using different types of networks at different numbers of epochs. By applying CNN model on test data. The classification results of different models are compared. Result comparison with different batch sizes used for tunning the DL [24]models, the accuracy of all networks is determined by different batch sizes 8,16,32. Batch sizes for all networks provides the highest performance of results so, size 32 is useful for accuracy.

Different Batch sizes by using CNN models testing accuracy in percentage.

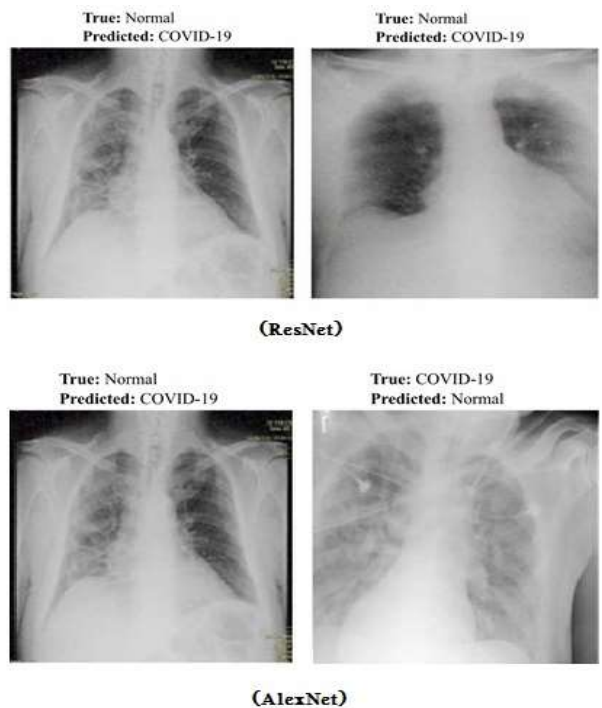

Fig.1 Output of models which shows the best results for COVID-19 detection

### B. Comparison the State-of-the Art Methods

When the CNN model's result is matched to the DL technique for the detection of coronavirus disease by utilising Chest X-Ray pictures, it's clear that the suggested model produced more fine-tuned results than the other methods. As compare to the studies [18], [19], [4] the suggested study considered an equitably large amount of specimens to confirm the CNN models while [24], [25]utilised similar but larger datasets to support their algorithms, however these datasets suffered from class variance issue and had a lower number of coronavirus infected cases in general, whereas the dataset used in this work had equal class distribution for healthy and coronavirus infected individuals. In paper [24] the issue of category variance was resolved by the use of offline amplification techniques, and it was also discovered that binary distribution was used in the majority of the research nevertheless, multiclass (mostly 3,4) grouping was performed in [24], [25]. Moreover, in the context of CoV class susceptibility, an efficiency distinction was made in between recommended approach and the current technique. The COVID-19 category suscepitvity of researchers [25] was the lowest, at 87.10 percent. It should be observed that suggested technique likewise obtained 100% CoV category suscepitvity [18], researcher [24] also obtained a 100 percent COVID-19 category suscepitvity. The primary test set employed in these research, on the other hand, has a similar lower number of coronavirus disease images.

TABLE II
COMPARISON WITH STATE-OF-ART DEEP LEARNING COVID-19 FINDINGS USING CHEST X-RAY IMAGES

| Method | Classes | Number of X-ray samples | Accuracy (%) |
|---|---|---|---|
| COVIDX-Net | CoV | 50 | 90.00 |
|  | Healthy | CoV: 25 and Norm: 25 |  |
| ResNet-50 | CoV | 100 | 98.00 |
|  | Healthy | COV: 50 & NORM: 50 |  |
| ResNet-50 and SVM | CoV | 50 | 95.38 |
|  | Healthy | COV: 25 & NORM: 25 |  |
| SqueezeNet and MobileNetV2 SMO and SVM | CoV | 458 | 98.25 |
|  | Healthy Pneumonia | COV: 295, NORM: 65 & PNEU: 98 |  |
| COVID-Net | CoV | 13800 | 92.60 |
|  | Healthy Pneumonia | COV: 183, NORM: – & PNEU: – |  |
| Bayes-SqueezeNet | CoV | 5949 | 98.30 |
|  | Healthy Pneumonia | COV: 76, NORM: 1583 & PNEU: 4290 |  |
| COVID-ResNet | CoV | 5941 | 96.23 |
|  | Healthy l Bacterial pneumonia Viral pneumonia | COV: 68 & NORM: – |  |
| ResNet-34 | CoV | 406 | 98.33 |
|  | Healthy | COV: 203 & NORM: 203 |  |

*C. Discussion*

We tested the efficacy of the most powerful CNN models for detecting CoV infection in X-Ray pictures, including ResNet-34, AlexNet, VGG-16, SqueezeNet, and MobileNet-V2. Larger tests were carried out on an oppositely big dataset, taking into account a variety of parameters in order to determine the best performing technique for automated coronavirus screening. The fact that this suggested model is verified using a very few coronavirus affected images is a serious flaw. There is currently no big dataset publically accessible; this is an ongoing and novel pandemic; nevertheless, we want to test our technique with larger datasets in the future.

The limitation of our research is that we observe some deep learning and machine learning models on different datasets, which can be limited. We observe the performance of different models on some classes which are related to text only.

## IV. CONCLUSIONS

Using Chest X-Ray pictures, this deep learning automated model was presented for effective utilization of COVID-19 affected individuals from normal instances. ResNet-34 outperformed the other competitor models with a precision of 98.3 percent, indicating that it may be used to predict COVID-19 infection. As a result, the efficacy of the suggested model for multi classification will be demonstrated in future research. We also aim to investigate the usage of optimization techniques in conjunction with the DarkCovidNet model to create a more authentic model for properly detecting infected individuals.

## ACKNOWLEDGMENT

We say thanks to ALLAH Almighty for everything throughout our research work. He is always with us for help us out of challenging moments of our life. We are grateful to ALLAH Almighty who made us able to complete our research work. It is due to HIS unending mercy that this work moved towards success. ALLAH Almighty bestowed us the opportunity and spirit to make the material assessments in already existing ocean of knowledge of the subject. All respect for his "HOLY PROPHET" (peace be upon him) who enabled us to recognize the purpose of creation.